\documentstyle[twoside,fleqn,espcrc2,epsfig]{article}

\def\tan2thc{{\rm tan}^2\theta_C}

\def\thanks#1{}

\newcommand{\AmS}{{\protect\the\textfont2
  A\kern-.1667em\lower.5ex\hbox{M}\kern-.125emS}}
\newcommand{\mev}{MeV/$c^2$ }
\newcommand{\gev}{GeV/$c^2$ }
\newcommand{\mypreprint}{\vskip -2cm
\hfill\hbox{{\vbox{\hbox{\hfil  OHSTPY-HEP-T-98-025 }
\hbox{\hfil DOE/ER/01545-756 }}}}
\vskip 2cm}

\title{Limits on the Mass of $\nu_\tau$ from CLEO 
\thanks{Invited talk at {\it The Fifth Workshop on Tau Lepton 
        Physics,} Santander, Spain, 14-18 September 1998.} }
\author{Jean E. Duboscq \address{Department of Physics, 
The Ohio State University, Columbus, OH 43210-1168   USA \\
email: jed@mail.lns.cornell.edu } 
\mypreprint }
\begin{document}

\begin{abstract}
A limit
 on the mass of the tau neutrino $m_{\nu_\tau}$ is derived from  
$4.5 \times 10^6$ tau pairs produced in an integrated luminosity of 
  5.0 $fb^{-1}$ of  
$e^+e^- \to \gamma* \to
 \tau^+ \tau^-$ reactions at center of mass energies $\approx$ 10.6 GeV. 
 The measurement technique involves a two-dimensional extended likelihood
  analysis, including  the dependence of the end-point population
 on $m_{\nu_\tau}$, and allows for the first time an explicit
 background contribution. We use 
 the decays $ \tau \to 5 \pi \nu_\tau$ and $\tau \to 3 \pi 2\pi^0 \nu_\tau$ 
 to obtain  an upper  limit of 30 \mev at $95\%$ C.L., as well as
 a preliminary limit of 31 \mev from $\tau \to 3 \pi \pi^0 \nu_\tau$
\end{abstract}


\maketitle

\section{Introduction}
We present two separate analyses of the mass of the $\nu_\tau$. 
The first analysis, as published in ~\cite{CLEO},  herein referred to
as  the $5h$ analysis, comprises a sample of  the decays 
 $ \tau \to 5 \pi \nu_\tau$ and $\tau \to 3 \pi 2\pi^0 \nu_\tau$. These decay
modes  have a small overall branching fraction but a significant probability 
 of populating the hadronic mass end-point sensitive to $m_{\nu}$. The
 second (preliminary) analysis, herein referred to as the $4h$ analysis, exploits
the decay $\tau \to 3 \pi \pi^0 \nu_\tau$. Although the probability per event
 of being near the endpoint is much smaller in this sample, the much larger
 branching ratio makes it competitive. 
\section{The CLEO Analysis Method}
\begin{figure}[htb]
\centerline{
\epsfig{file=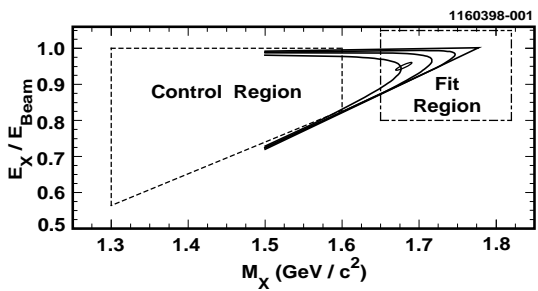,width=0.45\textwidth }}
\caption{ The number of events in the 
 the Fit Region  relative to the number in the
  Control Region
  in the scaled hadronic energy versus hadronic
 mass plane  is  a function of neutrino mass.
 Kinematically allowed neutrino mass contours are 
 drawn for neutrino masses of 0, 30, 60 and 100 \mev. 
 Note the typical error
 ellipse drawn in the fit region. }
\label{fig:fitregion}
\end{figure}

The analyses use the two-dimensional hadronic energy versus hadronic
 mass spectrum, first used in ~\cite{OPAL}, to obtain a limit on the mass
 of the $\nu_\tau$. The selection criteria require well reconstructed tracks,
 as well as good quality $\pi^0$ candidates, when appropriate, recoiling
 against a well separated one prong tag. For the $5h$ analysis, the
 tag track is required to be an identified lepton, while in the
 $4h$ analysis, the tag  must be consistent with either
 an electron, muon, $\pi$ or $\rho$. Stringent excess
 energy criteria reduce the possible backgrounds in both analyses.
 In the $5h$ analysis only a small region near the endpoint is used
 for the fit. For that analysis, an extra Poisson coefficient is used in an
 extended likelihood,  relating the number of 
 events expected in the Fit Region at the endpoint as a function 
 of neutrino mass
 to the number of events observed in data in a neutrino-mass-insensitive
 Control Region (see Fig~\ref{fig:fitregion}.)
 The fit region for the $4h$ analysis is chosen 
 large enough to make the Poisson term unnecessary.
 For the first time in any $\nu_\tau$ mass analysis , both the $5h$ and
 $4h$ fits allow for an explicit background term to account for non-signal 
 events.
\section{The $5h$ Sample }
\begin{figure}[htb]
\centerline{
\epsfig{file=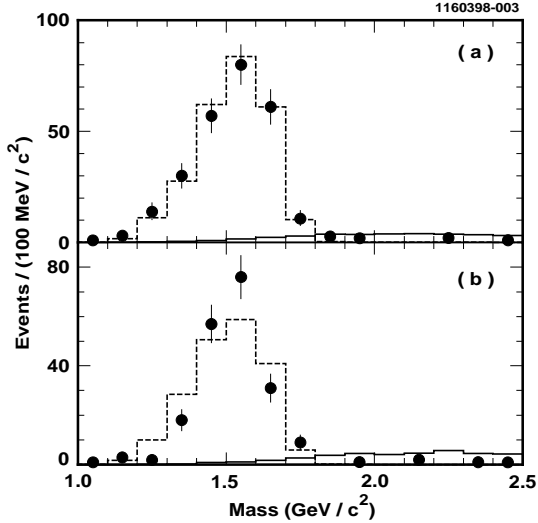,width=0.45\textwidth }}
\caption{ The hadronic mass for the $5\pi$ (a) and $3\pi2\pi^0$ (b) event
 candidates. The Monte Carlo estimation of signal shape, normalized to 
 the data sample size in the control region, is displayed as the dotted
 histogram. The solid histogram is the background estimator displayed at
 five times its true size for illustration purposes.}
\label{fig:fivemass}
\end{figure}

\begin{figure}[htb]
\centerline{
\epsfig{file=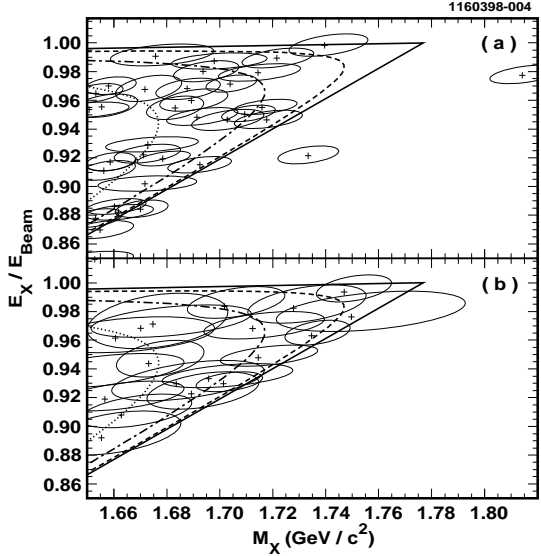,width=0.45\textwidth }}
\caption{ The hadronic energy (scaled to beam energy) vs 
 hadronic mass for (a) the $5\pi$ and the (b) $3\pi2\pi^0$  event
 candidates in the fit region. Ellipses represent 68$\%$ confidence
 levels. }
\label{fig:fiveevsm}
\end{figure}

The final event sample is summarized in Table~\ref{tab:fivehsummary}. The
 hadronic masses of the $5h$ samples are displayed in Fig~\ref{fig:fivemass}.
 There are  207 events in the $3\pi 2\pi^0$ sample and  
 266 events in the $5\pi$ sample. Backgrounds can come from either
 misreconstructed tau decays or from non-tau sources such as charm decays. 
 The non-tau background estimator shape used in
 the fit is obtained from a data sample of events with tracks in
 one hemisphere with loose selection criteria and invariant mass above the
 tau mass recoiling against  either $5\pi$ or $3\pi2\pi^0$ candidates. 
 Its normalization is
 provided by the number of events in Fig~\ref{fig:fivemass} above the tau
 mass\footnote{ Corrections for $\tau$ event feed-across into the background
 estimator are negligible.}. Backgrounds from tau decays to final
 states other than $5\pi$ and $3\pi2\pi^0$ are estimated from Monte Carlo
 studies\footnote{ The CLEO Monte Carlo is based on 
 ~\cite{lund,geant,koralb,photos}. }, and are expected to be small. 
\begin{table*}[hbt]
\setlength{\tabcolsep}{1.5pc}
\newlength{\digitwidth} \settowidth{\digitwidth}{\rm 0}
\catcode`?=\active \def?{\kern\digitwidth}
\caption{Summary of $5h$ and $4h$  analyses }
\label{tab:fivehsummary}
\begin{tabular*}{\textwidth}{lrrr}
\hline
\hline
   Mode     & $5\pi$ & $3\pi2\pi^0$  & $4h$ \\ \hline
Total events                &  266      & 207 & $29 \times 10^3$ \\
Events in Fit Region         &  36& 19  & $17 \times 10^3$ \\ 
Fit Region Purity ($\%$)     & 99 &  93 & 90 \\ 
Selection Efficiency ($\%$)  &   3.1& 0.4 & 2.6  \\
Typical Mass Resolution (\mev) &   15 &  25 & 20 \\ 
Typical Energy Resolution (MeV) & 25 &   50 & 38 \\ 
Uncorrected Upper Limit $@$ 95 $\%$ CL  (\mev) &  31 &  33 & 26 \\ 
\hline
\end{tabular*}
\end{table*}
 The hadronic energy scaled to beam energy versus hadronic mass distribution
 is shown in Fig~\ref{fig:fiveevsm} for events in the neutrino mass
 sensitive Fit Region. 
\section{The $4h$ Sample }
\begin{figure}[htb]
\centerline{
\epsfig{file=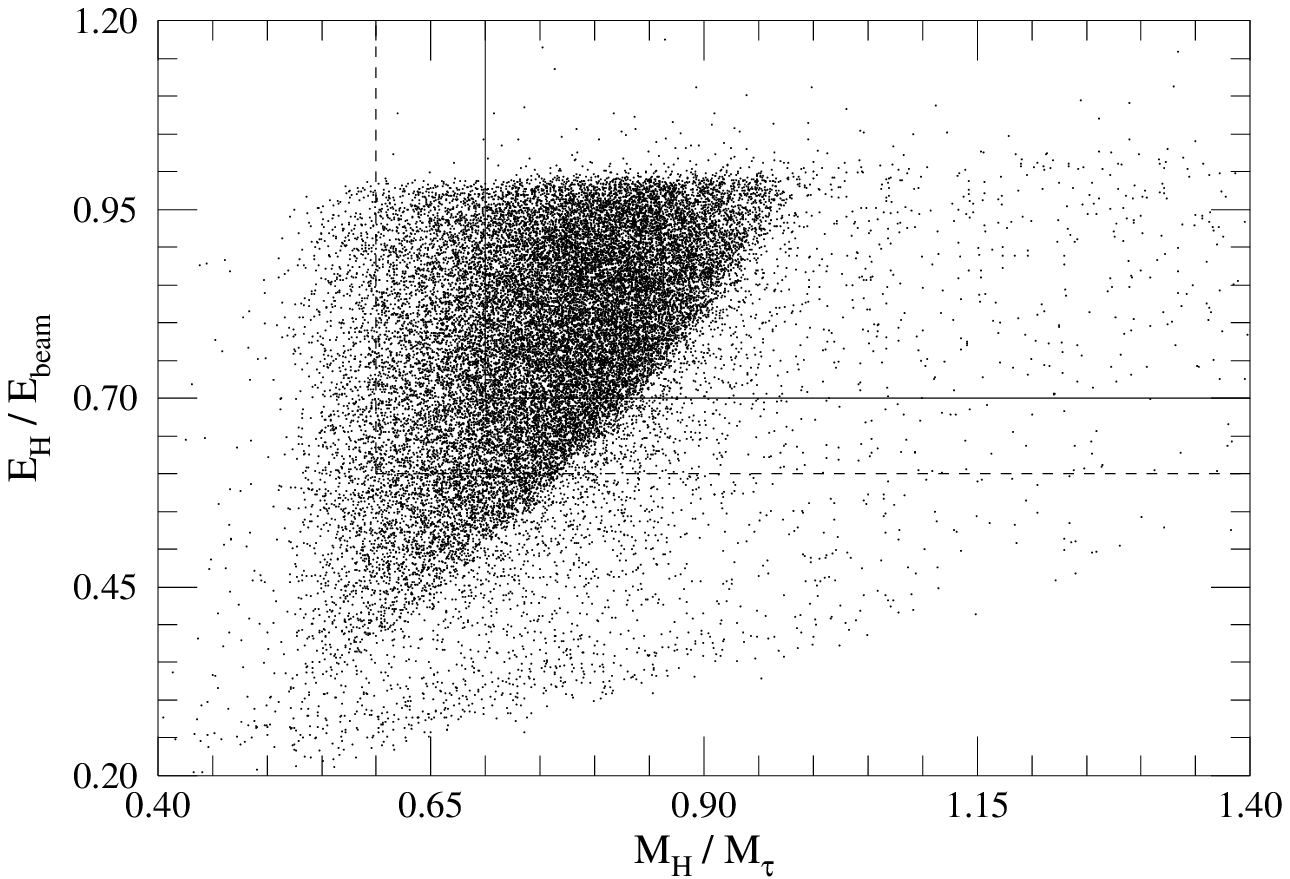,width=0.5\textwidth }}
\caption{ The hadronic energy (scaled to beam energy) vs 
 hadronic mass (scaled to $\tau$ mass ) for the $4h$ sample. The solid lines
 indicate the Fit Region, while the dashed line indicates a larger
 fit region used in systematics studies.  }
\label{fig:fourevsm}
\end{figure}

 The scaled energy versus mass distribution of the full sample of $4h$ decays
 is shown in Fig~\ref{fig:fourevsm}. As noted in table~\ref{tab:fivehsummary},
 there about 29K  events in the sample, with 17K  events in the fit region.
 In addition to the dominant $3\pi\pi^0$ final state, the data sample includes 
 other $3h \pi^0$ resulting from the $K_s \pi\pi^0$, $K\omega$, 
 $K K_s\pi^0$ and $\pi \omega \pi^0$ intermediate hadronic states. All these 
 are  included in the fit. The backgrounds are predominantly from
 misreconstructed tau events in the Fit Region ($7\%$) and decays of
 $q\overline{q}$ pairs ($3\%$). These background shapes are estimated
 from Monte Carlo. The tau background sample normalization is estimated
 from Monte Carlo, and the $q\overline{q}$ background normalized to the number
 of events seen in the data above the tau mass, corrected for $\tau$ 
 feedthroughs.
\section{ Spectral Functions }
\begin{figure}[htb]
\centerline{
\epsfig{file=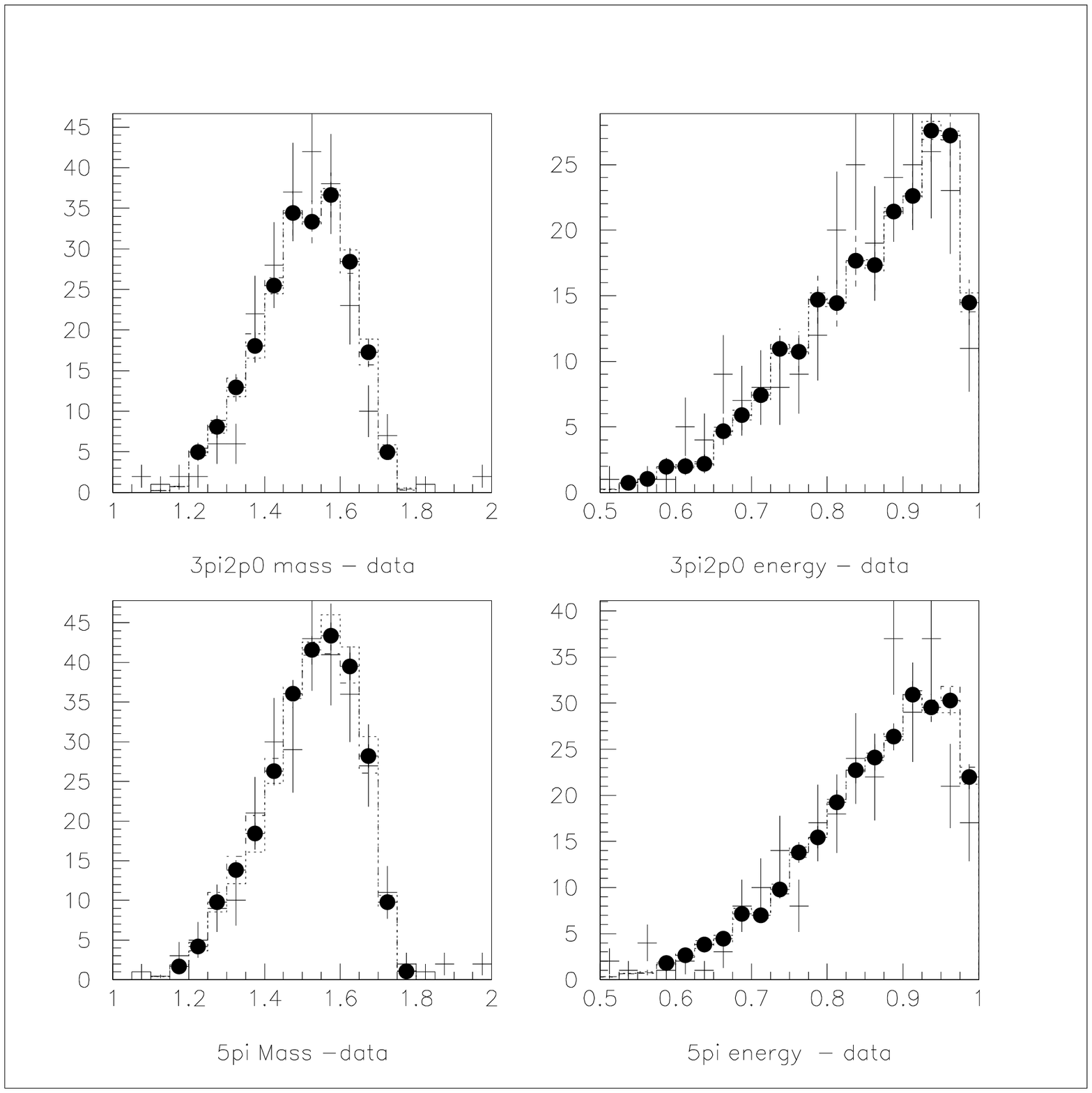,width=0.45\textwidth,
bbllx=23pt, bblly=148pt, bburx=539pt, bbury=672pt,clip= }}
\caption{ The fitted spectral functions compared to the data
  for the $5h$ samples. The top
 (bottom)  plots show the $3\pi2\pi^0$ ($5\pi$) 
 sample mass fit and resulting energy  projections. The crosses represent
 the $5h$ data, while the dots represent the fit, while the estimated errors
 in the fit are shown by the histograms. Note that the fit was performed
 on a statistically independent sample.  }
\label{fig:fivehspectral}
\end{figure}

\begin{figure}[htb]
\centerline{
\epsfig{file=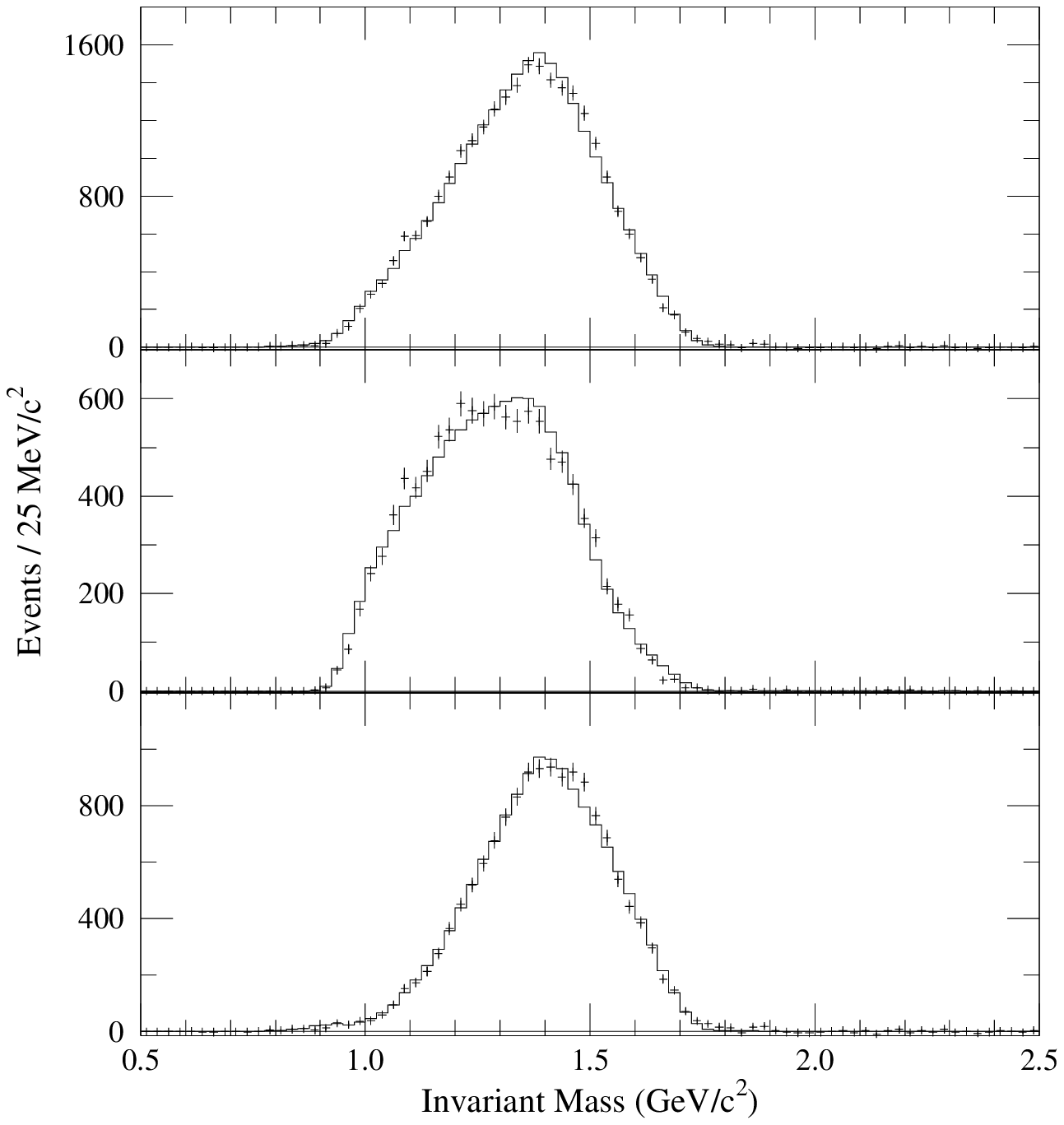,width=0.45\textwidth }}
\caption{ The spectral function used in the $4h$ sample fit, for events
 containing no $\omega$ (top), those containing an $\omega$ (middle), 
 and all events (bottom).}
\label{fig:fourhspectral}
\end{figure}

 The invariant mass spectra used in the $5h$  analysis are fit in an
 independent data sample of $\pi$ tagged $5h$ decays. The fit functions
 are derived from that expected from $e^+e^- \to 4\pi$ distributions, along with
 soft pion theorems, as well as the shape expected for 
 $\tau \to 6\pi\nu$, as extrapolated from $e^+e^- \to 6 \pi$ 
 distributions~\cite{Tsai,Gilman,Pham}. 
 The former shape is expected to dominate, while the latter shape is allowed
 in the fit as a purely phenomenological tuning parameter. Since the $6\pi$ shape
 expects events at higher masses than the $4\pi$ shape, its inclusion naturally
 leads to a more conservative limit. The fit also includes a background shape,
 and excludes the mass range within 100 \mev of the tau mass to decrease bias on the neutrino
 mass. The final distributions are shown in Fig~\ref{fig:fivehspectral}.
The $4h$ spectral function is dominated by the $3\pi\pi^0$ final state.
 This state arises via intermediate decays including a
 $\rho$, $\rho'$, $\rho ''$ or an $\omega$. The spectral function is
 obtained  by the fitting  all $2\pi$, $3\pi$ and $4\pi$ combinations  
 in the current to   a sum
 of such intermediate states, as an extension of the model used in
 Tauola\cite{koralb}:
\begin{eqnarray*}
J= & & F^{\rho\pi\pi}(Q_{4\pi}^2)\Sigma_{i=1,5}{\cal A}_i f^\rho_i(q_{2 i}) \\
 & + & F^{\omega\pi}(Q_{4\pi}^2)\Sigma_{i=1,2}{\cal A}_i f^\omega_i(q_{3 i}) 
\end{eqnarray*}
 where $F^{\rho\pi\pi}$, $F^{\omega\pi}$, and $f^\rho $ are sums of
 Breit-Wigner distributions for the $\rho$, $\rho'$ and $\rho''$,
 and $f^\omega$ is a Breit-Wigner distribution for the $\omega$.
 The $4\pi$ combinations used are only those with a mass below
 1.6 \gev. Other $3h\pi^0$ decays modes not considered in the
 above are small and fixed to their Monte Carlo expected shapes. 
 Although this is not a unique model for this decay, all mass
 distribution projections fit very well - see Fig~\ref{fig:fourhspectral}.
\section{ The Likelihood  Functions  }
The event likelihood includes both
 a signal shape with a dependence on neutrino mass, and a background shape:
\begin{eqnarray*}
{\cal L}_{event}(M_\nu) = & & \alpha  {\cal L}_{sig}(data, M_\nu)  \\
   & + & (1-\alpha) {\cal L}_{bgd}(data)
\end{eqnarray*}
where $\alpha$ sets the relative background  and signal 
sizes\footnote{ Note that  an a priori assumption on the neutrino mass allowed
 range must be made for normalization purposes. In this case we chose to assume
 that  $M_\nu < 100 \,\,{\rm MeV/}c^2$. The result is not strongly dependent on this
 assumption for reasonable a priori distributions. }.  
 The full likelihood is a product over the individual
 event likelihoods, and, for the $5h$ analysis, includes a 
  Poisson coefficient $P(N_{obs},M_\nu) $ to account for the expected
 number of events in the Fit Region given the number of events in the
 Control Region, and the neutrino mass. 

 The signal likelihood used in previous analyses~\cite{LUCA} relied
 on convolutions of parametrizations of Monte Carlo data. The present
 analyses forego the unnecessary convolution and parametrisation steps, 
 and fit directly to  Monte Carlo data. Reconstructed Monte Carlo events, 
 generated with
 a massless neutrino, are accepted or rejected by the data selection
 criteria. Their generated mass and energy are then  smeared according to 
 an analytic detector smearing
 function, depending only on hadronic mass and energy, and their
 associated propagated tracking and calorimetry errors. 
 Furthermore, this smearing function is taken as universal
 for all events in the same sample when scaled according to propagated mass 
 and energy errors. The probability that a  Monte Carlo event could
 be reconstructed as the data event in question is proportional to
 the smearing. Symbolically,
\begin{equation}
{\cal L}_{sig} \propto \sum_{MC} G( {\tilde{X}_{data}-X_{MC} \over{\sigma}} ) 
  {\cal W}(X_{MC}, M_\nu) 
\end{equation}
where $G$ represents the detector smearing function, and ${\cal W}$ is an
 exactly calculable weight factor for neutrino mass hypotheses relative
 to a massless neutrino.
 The smearing function is obtained from Monte Carlo $\tau$ decays and
 is parametrised as the sum of three 2 dimensional Gaussians.
\section{Results }
\begin{figure}[htb]
\centerline{
\epsfig{file=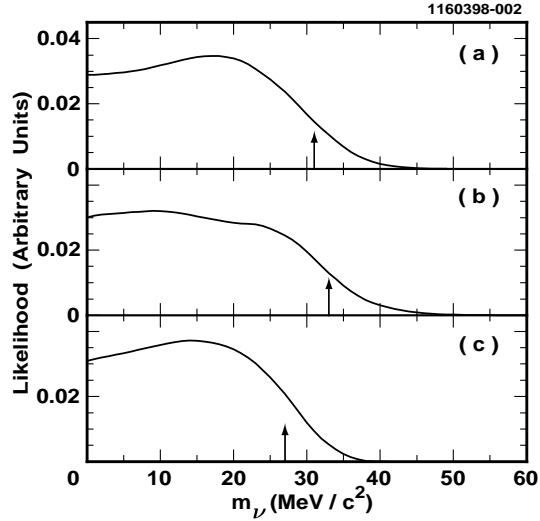,width=0.45\textwidth , clip=}}
\caption{ The likelihood  function versus neutrino mass 
from the  $5h$ sample fit for the $3\pi2\pi^0$ sample (top),
 $5\pi$ sample (middle) and the combined $5h$ sample (bottom).}
\label{fig:fivehlikeli}
\end{figure}

\begin{figure}[htb]
\centerline{
\epsfig{file=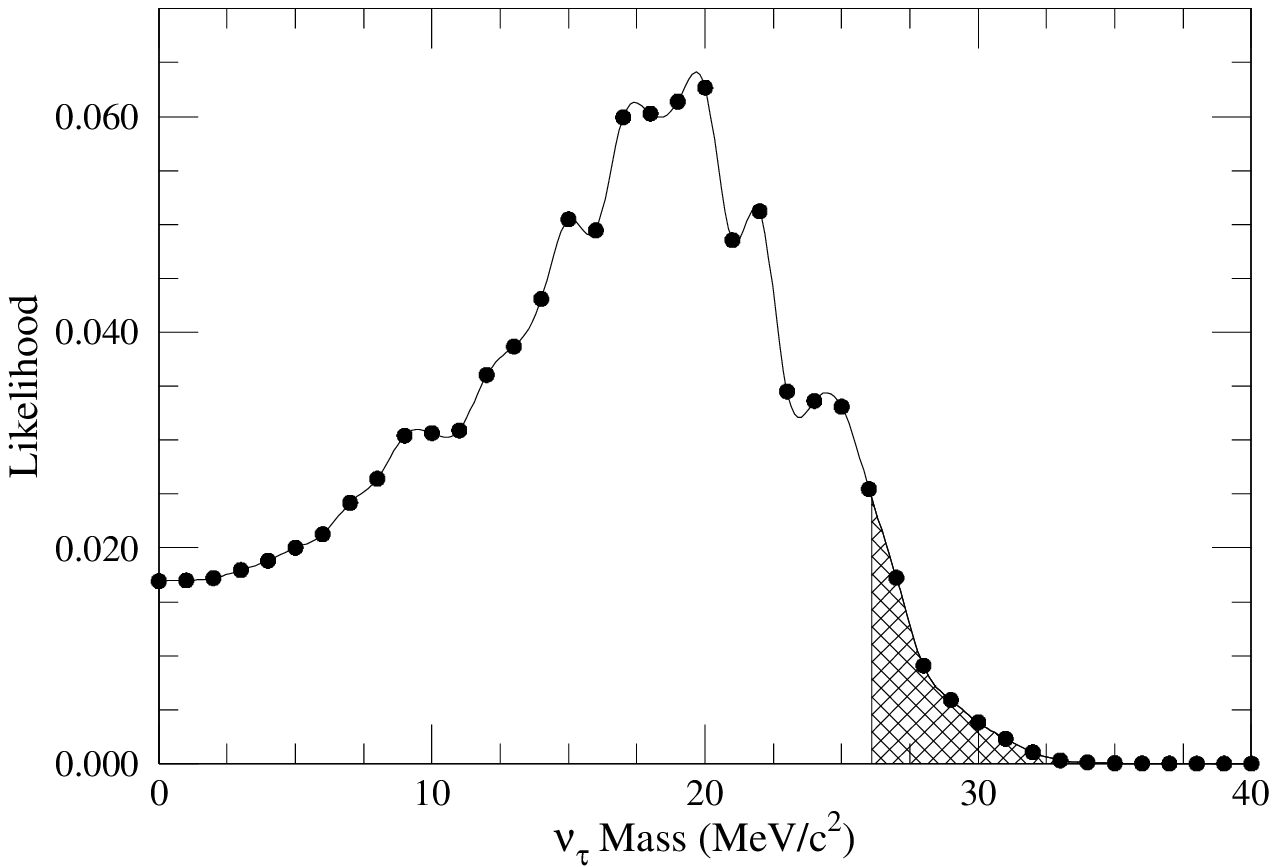,width=0.45\textwidth,clip=}}
\caption{ The likelihood  function versus neutrino mass 
from the  $4h$ sample fit. The small bin to bin variations result from 
 limited Monte Carlo statistics, and are not significant.}
\label{fig:fourhlikeli}
\end{figure}

 The final likelihoods are shown in Fig~\ref{fig:fivehlikeli} and
 Fig~\ref{fig:fourhlikeli}. An integral of the likelihood from  0 \mev
 to $95\%$ of the full area of the likelihood gives raw upper limits
 of 33, 31 and 26 \mev  respectively for the $3\pi2\pi^0$, $5\pi$ and
 $4h$ samples. Combining the  $3\pi2\pi^0$ and  $5\pi$ likelihoods
 give a raw upper limit of 27 \mev. Including  the $4h$ likelihood
 lowers the uncorrected upper limit to 25 \mev. 
 Systematic errors were carefully examined for the combined  $5h$  studies and
 the $4h$ study. 
 Charm meson decays to $K\pi$, $K2\pi$ and $K3\pi$ final states in 
 data and Monte Carlo
 show good agreement for the smearing functions, and are used to estimate
 systematic errors for the smearing function along the mass axis.
  Since $B$ mesons at CLEO are produced at rest, the mass 
 spectra of hadronic final states of $B$ meson decays give a direct measure 
 of the smearing along the energy axis. Smearing modeling systematics
  contribute 1.5 and 0.4 \mev to the $5h$ and $4h$ studies respectively.
 The spectral functions are all
 allowed to vary according to the uncertainties in their associated fits, 
 contributing 1.9 and 1.2 \mev respectively.
 The CLEO mass and momentum reconstruction scales are also considered
 as sources of error (1.5 and 2.3 \mev  respectively) as was the
 CLEO energy scale (0.2 and 3.7 \mev)
\footnote{ Differences in nomenclature
 for systematic errors between the studies explain the apparent size
 inconsistencies of systematic errors. For instance, modeling of the
 $\pi^0$ contribution in the smearing is accounted for as a smearing
 systematic in the $5h$ study, and as a momentum scale systematic error 
 in the $4h$ study. } The sum in quadrature of all considered 
 systematic errors is 3.1 \mev for the
 $5h$ study and 5.1 \mev for the $4h$ study. Following the conservative
 prescription used by the LEP experiments~\cite{LUCA}, these errors are added linearly
 to the upper limits, resulting in an upper limit of 30 \mev for the
 $5h$ sample and a preliminary limit of 31 \mev  for the $4h$ sample,
 all at the $95\%$ confidence level. The combination of the $5h$ and $4h$
 studies is still under consideration at this writing.
\begin{table}[]
\setlength{\tabcolsep}{1.0pc}
\begin{tabular*}{0.5\textwidth}{lrr}
\hline
 $M_\nu^{in}$ \mev   & 25 events      & 450 events \\
\hline
 0       & $3\%$          & $67\%$ \\
 50      & $\approx 1\%$  & $<1 \%$ \\
\hline
\end{tabular*}
\caption{Probability of a 95th percentile below 27 \mev
 for small and large data samples given a true neutrino mass 
 of  0 and 50 \mev.}
\label{tab:badprob}
\end{table}
 The interpretation of the 95th percentile of a likelihood as a statement
 of probability about the neutrino mass is dependent upon the experimentally
 reconstructed 
 likelihood being representative of an ensemble of likelihoods for samples
 drawn from a distribution with similar properties. Thus conclusions drawn
 from experiments with small statistics must be interpreted with care 
 especially
 when  very stringent limits are quoted. These limits tend to have much lower
 discriminatory power than one might naively think.
 Table~\ref{tab:badprob} illustrates
 this for studies done with the CLEO Monte Carlo. For a sample of 25 events,
 a 95th percentile limit of 27 \mev is obtainable $3\%$ of the time 
 with a massless neutrino
 but it is also obtainable about $1\%$ of the time with a 50 \mev
 neutrino. The discriminatory power is of course much better for 
 larger samples. Extensive Monte Carlo studies have shown that the 
 95th percentile upper limits obtained for the $5h$ samples in this
 study are not unlikely - in some 35 $\%$ (74 $\%$) of Monte Carlo experiments
 with similar statistics, the 95th percentile was found to be larger
 than that found in the data 
 for the $5\pi$ ( $3\pi2\pi^0$ ) sample.  Thus the interpretation of
 the combined 95th percentile as  an upper limit is credible.
 
 In all the  fits presented here, the likelihood peaks away from 0 \mev.
 In no case however is the peaking deemed to be significant. In the 
 $4h$  study, the ratio of likelihoods at the peak relative to that
 at 0  \mev corresponds to a 1.6 $\sigma$ effect if one assumes that
 the underlying probability densities are Gaussian. Multiple Monte Carlo
 experiments, as well as basic statistical theory,  also show that the 
 likelihood function is not always expected to peak at 0  \mev, even for
 a truely massless neutrino. 
 In addition the conservative approximations made
 in the analysis can not only result in less stringent limits, but can also
 have the effect of pushing any peak further from  0 \mev. These likelihood
 curves should not therefore be used to infer the existence of a massive
 neutrino.
\section{Conclusions}
Using the world's largest dataset,  CLEO sets a combined upper limit
 of 30 \mev for a combined study of  $\tau \to 5\pi \nu$, $\tau \to 3\pi 
 2\pi^0 \nu$ decays and a preliminary upper limit of 31 \mev from the
 decay $\tau \to 3h \nu$. These limits use a novel Monte Carlo integration
 technique, and are the first to explicitly account for possible
 backgrounds in the likelihood function.

\end{document}